\documentclass[epj]{webofc}
\usepackage[varg]{txfonts}   
\woctitle{Hot Planets and Cool Stars}
\begin{document}
\title{The Next Generation Transit Survey (NGTS)}
%
%

\author{
Peter J.\ Wheatley\inst{1}\fnsep\thanks{\email{P.J.Wheatley@warwick.ac.uk}} 
\and
Don L.\ Pollacco\inst{1}
\and
Didier Queloz\inst{2}
\and
Heike Rauer\inst{3,4}
\and
Christopher A.\ Watson\inst{5}
\and
Richard G.\ West\inst{6}
\and
Bruno Chazelas\inst{2}
\and
Tom M.\ Louden\inst{1}
\and
Simon Walker\inst{1}
\and
Nigel Bannister\inst{6}
\and
Joao Bento\inst{1}
\and
Matthew Burleigh\inst{6}
\and
Juan Cabrera\inst{3}
\and
Philipp Eigm\"{u}ller\inst{3}
\and
Anders Erikson\inst{3}
\and
Ludovic Genolet\inst{2}
\and
Michael Goad\inst{6}
\and
Andrew Grange\inst{6}
\and
Andr\'{e}s Jord\'{a}n\inst{7}
\and
Katherine Lawrie\inst{6}
\and
James McCormac\inst{8}
\and
Marion Neveu\inst{2}
}

\institute{Department of Physics, University of Warwick, Coventry CV4 7AL, UK
\and
Observatoire Astronomique de l’Universit\'{e} de Gen\`{e}ve, 
1290 Sauverny, Switzerland
\and
Institut f\"{u}r Planetenforschung, Deutsches Zentrum f\"{u}r Luft- und Raumfahrt, 
12489 Berlin, Germany 
\and
Zentrum f\"{u}r Astronomie und Astrophysik, Technische Universit\"{a}t Berlin, 
10623 Berlin, Germany
\and
Astrophysics Research Centre, Queen’s University Belfast, Belfast BT7 1NN, UK
\and
Department of Physics and Astronomy, University of Leicester, Leicester LE1 7RH, UK
\and
Departamento de Astronom\'{i}a y Astrof\'{i}sica, Pontificia Universidad Cat\'{o}lica de Chile, 
Santiago, Chile
\and
Isaac Newton Group of Telescopes, 
38700 Santa Cruz de la Palma, Canary Islands, Spain
          }

\abstract{%
The Next Generation Transit Survey (NGTS) is a new ground-based sky survey designed to find transiting Neptunes and super-Earths. By covering at least sixteen times the sky area of {\em Kepler}, we will find small planets
around stars that are sufficiently bright for radial velocity confirmation, mass determination and atmospheric characterisation. The NGTS instrument will consist of an array of twelve independently pointed 20\,cm telescopes fitted with red-sensitive CCD cameras. It will be constructed at the ESO Paranal Observatory, thereby benefiting from the very best photometric conditions as well as follow up synergy with the VLT and E-ELT. Our design has been verified through the operation of two prototype instruments, demonstrating white noise characteristics to sub-mmag photometric precision. Detailed simulations show that about thirty bright super-Earths and up to two hundred Neptunes could be discovered. Our science operations are due to begin in 2014. 
}
\maketitle
\section{Introduction}
\label{intro}
Ground-based transit surveys such as WASP \cite{Pollacco06} and HAT \cite{Bakos04} have discovered an extraordinary variety of mainly Jupiter and Saturn-sized exoplanets that are challenging our understanding of giant planet formation and migration. Discoveries have included planets that are variously highly-inflated, extremely close-in, or in retrograde orbits. Meanwhile, space-based transit surveys such as CoRoT \cite{Auvergne09} and {\em Kepler} \cite{Borucki10} have extended our sensitivity to smaller planets, finding the first rocky exoplanets as well as a startling variety of multiple systems. However, due to the relatively narrow fields of view of space-based surveys, most small planet candidates orbit stars that are too faint for radial-velocity confirmation. As a consequence, their masses are often unknown or poorly known, placing only weak constraints on their bulk composition. Most of these planets are also too faint for atmospheric characterisation, even with future instrumentation such as the E-ELT, JWST, EChO or FINESSE. There is thus a strong scientific imperative to find small exoplanets around brighter stars. 

Building on experience gained in the WASP project, we have designed the Next Generation Transit Survey (NGTS) with the primary aim of discovering transiting Neptunes and Super-Earths around bright stars from the ground. 
Our 
objective is to find a statistically significant sample of such systems that are bright enough for radial-velocity confirmation and hence mass and density determination. By constraining the bulk compositions of our sample we will test population synthesis models of planet formation and evolution \cite[e.g.][]{Mordasini12}. Our brightest Neptunes and Super-Earths will also be suitable for atmospheric characterisation by transmission spectroscopy and secondary eclipse observations. 

\section{The NGTS instrument}
Our scientific goals require the detection of transits with depths at the 1\,mmag level. While this level of accuracy is routinely achieved from the ground in narrow-field observations of individual objects, it is unprecedented for a ground-based wide-field survey.  To reach this level of accuracy we have drawn on experience from the WASP project in order to minimise known sources of red noise related to imprecise pointing, focus and flat fielding. 

The NGTS facility will be an array of twelve 20cm f/2.8 telescopes on independent equatorial mounts, each fitted with a red-optimised large-format CCD camera. The facility will be sited at the ESO Paranal Observatory (Chile) in order to benefit from the world’s best photometric conditions. There is also excellent synergy with ESO facilities, including HARPS and ESPRESSO for radial-velocity confirmation, SPHERE for imaging wider-separation companions, and a variety of VLT and planned E-ELT instruments for atmospheric characterisation.  

The telescopes are astrographs of a custom design by ASA in Austria. They have a carbon fibre tube, hyperbolic primary mirror and custom corrector optics that provide a uniform 
point spread function across the entire 3\,deg field of view (FWHM of 1.1 pixels). The telescopes are fitted with a 400\,mm baffle in order to minimise sensitivity to scattered moonlight. 
The CCD cameras are also of custom design, by Andor Technology in the UK, employing e2v CCDs with back-illuminated deep-depletion silicon and anti-fringing technology that is optimised for the far red of the optical spectrum. This improves our sensitivity to smaller stars (K and M types) and hence smaller planets. 

The telescopes and cameras are mounted on independent equatorial fork mounts 
by OMI in the USA, and will be located in a single enclosure with a slide 
off-roof by GR PRO in the UK. The enclosure is designed to shelter the telescope units from the prevailing Northerly wind, while allowing each unit to point independently without obscuring the view of any other. 
The full telescope array has an instantaneous field of view of 96 square degrees, and we intend to observe around four fields each year. 
The instrument design is described in more detail in \cite{Chazelas12}.

\begin{figure}
\centering
\includegraphics[height=5cm,clip]{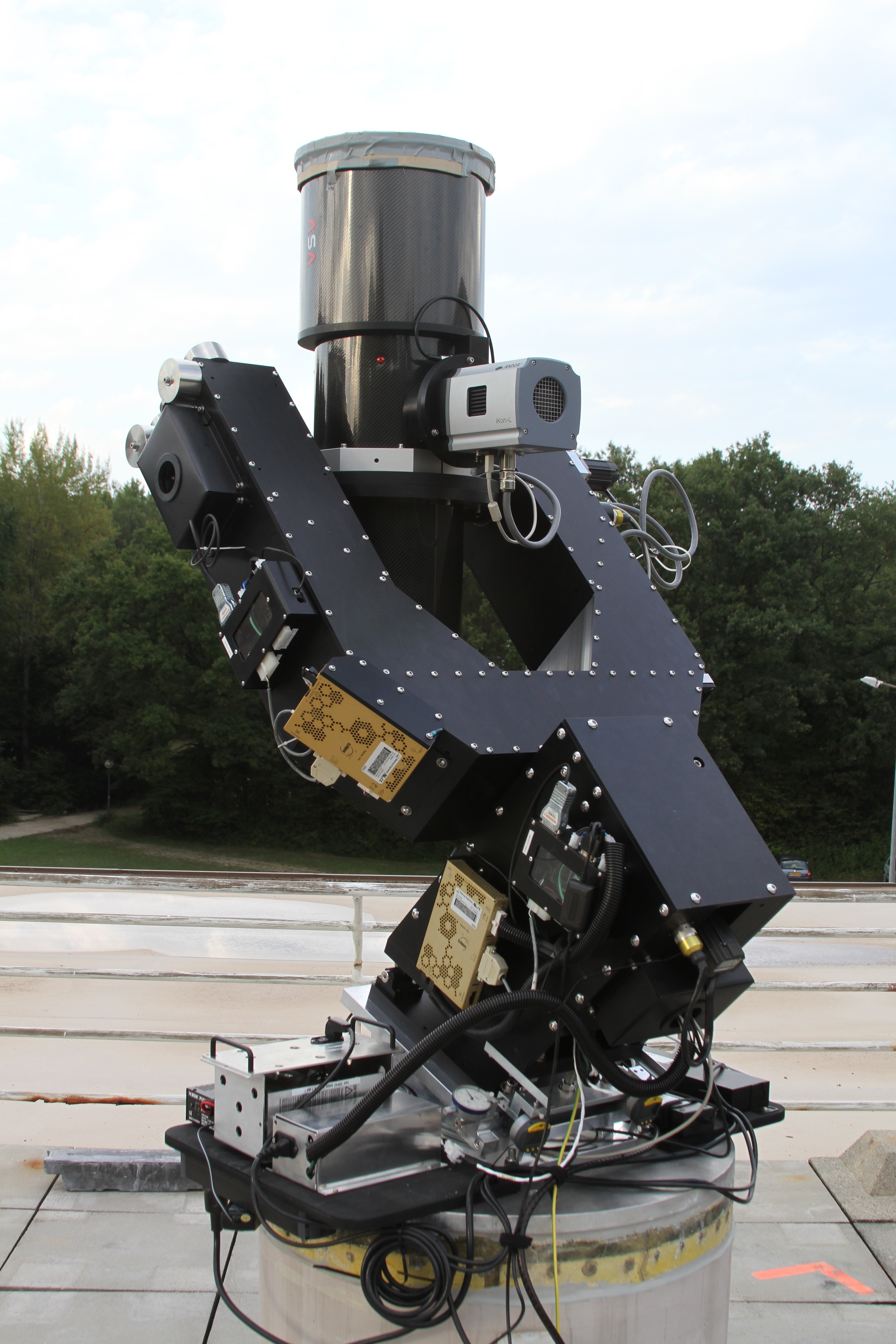}
\hspace{2mm}
\includegraphics[height=6cm,clip]{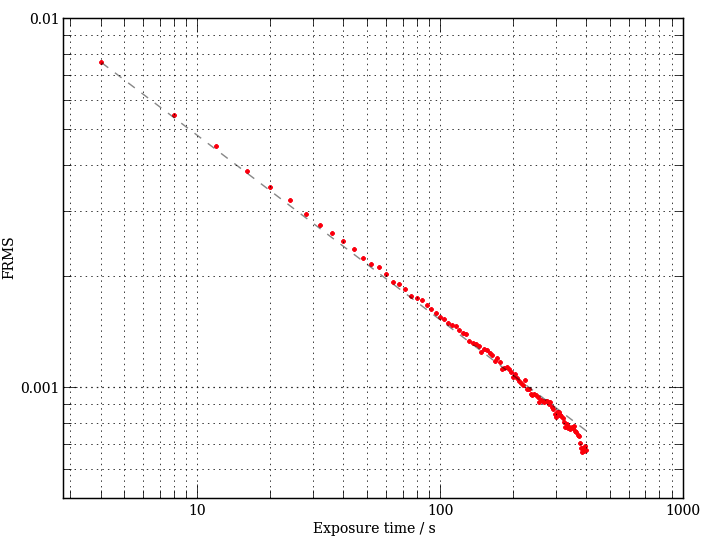}
\caption{{\bf Left:} the first of twelve NGTS telescope units. This unit was operated at Geneva during summer 2012. {\bf Right:} the noise characteristics of the NGTS telescope unit measured in Geneva. These noise measurements have been made using an ensemble of bright stars from across the whole field of view. The results show white noise behaviour to well below 1\,mmag precision. 
}
\label{fig-inst}       
\end{figure}

The deployment phase of NGTS is fully funded by our consortium institutes (DLR, Geneva, Leicester, QUB, Warwick) and most of the equipment has been procured. A complete telescope unit has been assembled and tested in Geneva, shown in the left hand panel of Fig.\,\ref{fig-inst}. On-site construction is due to begin in summer 2013, with science operations from 2014. Our data will be made publicly available after a proprietary period through the normal ESO archive. 

\section{The NGTS prototypes}
In order to develop and verify our instrument design we have deployed two prototype instruments. The first was operated on La Palma in 2010, and it verified the photometric performance of the back-illuminated deep-depletion e2v CCDs, and demonstrated the value of precise autoguiding. The second prototype was the complete NGTS telescope unit shown in Fig.\,\ref{fig-inst}. Tests with this unit have verified the optical performance of the telescope, as well as the pointing precision of the mount. The right hand panel of Fig.\,\ref{fig-inst} shows the results of end-to-end tests of our photometry from Geneva on the {\em Kepler} field. The median fractional RMS variability of an ensemble of bright stars from across the field of view is plotted as a function of binned exposure time. The measured noise is consistent with the expected scintillation level for the Geneva site \cite{Dravins98}. We initially found a noise floor of 3-mmag that we ascribe to variable water vapour extinction at the Geneva site, but since this 
was strongly correlated between all stars it was readily detrended using the standard SysRem algorithm \cite{Tamuz05}. Fig.\,\ref{fig-inst} then shows purely white-noise down to sub-mmag photometry.

\section{Simulated planet catch}
\begin{figure}
\centering
\sidecaption
\includegraphics[width=8.8cm,clip]{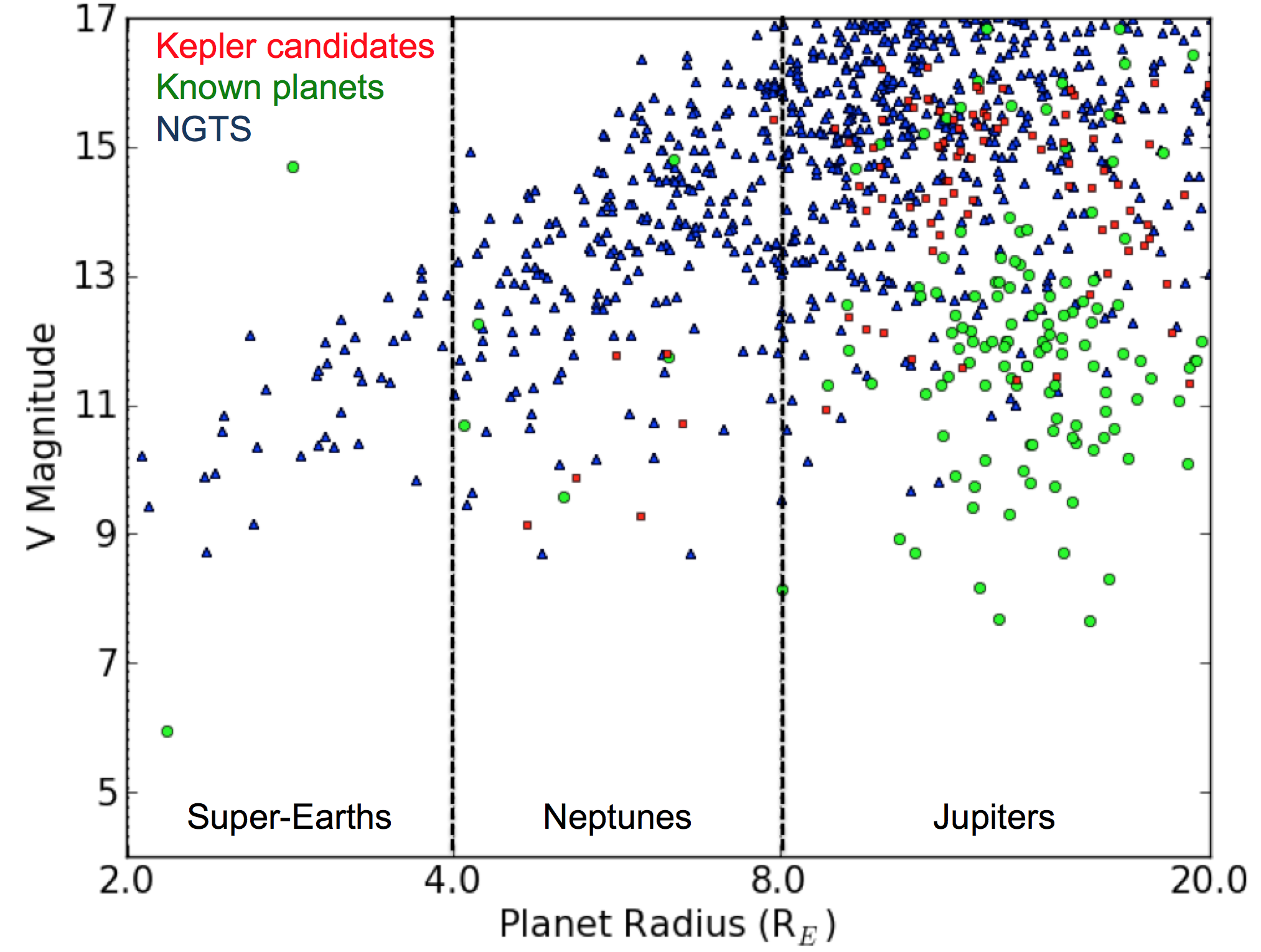}
\caption{Our simulated population of NGTS planets that can be confirmed in 10\,h with HARPS or ESPRESSO (blue). These are compared with the known transiting planets with radial-velocity confirmation (green) and the {\em Kepler} candidates that are confirmable with HARPS-N (red). This simulation shows a total of 39 confirmable super-Earths from NGTS and 231 Neptunes. }
\label{fig-sim}       
\end{figure}

We have performed detailed simulations of the NGTS planet catch for a four year baseline survey that covers sixteen times the area of the {\em Kepler} field.  A population of host stars was drawn from the Besancon model of the Galaxy \cite{Robin04}, and assigned a population of planets based on the planet size distributions and occurrence rates from {\em Kepler} \cite{Howard12}. This population was then sampled with the known characteristics of the NGTS prototype instruments, accounting for realistic source and background spectra, weather statistics for Paranal and scintillation. The resulting planet candidate population was then filtered with the sensitivity limits of the HARPS and ESPRESSO radial-velocity spectrographs, and the final predicted confirmable population is shown in Fig.\,\ref{fig-sim} assuming a total of 10\,h exposure time per candidate. We find that HARPS/HARPS-N observations are capable of confirming 37 Neptunes from NGTS compared with only 7 from {\em Kepler}. 
Allowing the HARPS-N exposure times to increase to 20\,h per candidate retrieves 21 Neptunes from {\em Kepler}, but still only 1 super-Earth. In contrast, follow up of NGTS candidates with ESPRESSO on the VLT is sensitive to 231 Neptunes and 39 super-Earths (with only 10\,h total exposure time per candidate).

The smallest NGTS confirmed planets will be prime targets for the ESA S Mission CHEOPS, which will provide precise radii and hence densities of our 
super-Earths. 
As well as testing models of bulk composition of super-Earths, this will allow us to prioritise objects by scale height for atmospheric follow up with VLT and {\em HST}, and eventually E-ELT and {\em JWST} (as well as dedicated missions such as EChO or FINESSE).


 \bibliography{wheatley_astroph}
%
%
%
%

\end{document}